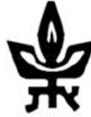

# Coherence Properties of Discrete Static Kinks

by

**Haggai Landa**

Thesis submitted towards the degree of M.Sc. in physics

Under the supervision of **Prof. Benni Reznik**

February, 2009

# Abstract

A chain of interacting particles subject also to a nonlinear on-site potential admits stable soliton-like configurations : static kinks. The linear normal-modes around such a kink contain a discrete set of localized, gap-separated modes. Quantization of the Hamiltonian in these modes results in an interacting system of phonons. We investigate numerically the coherence properties of such localized modes at low temperatures using a non-Markovian master equation. We show that low decoherence rates can be achieved in these nonlinear configurations for a surprisingly long time. If realized in the ion trap, kink internal modes may be advantageously used for Quantum Information Processing.

# Acknowledgements

I would like to express my gratitude to Prof. Benni Reznik for handing me this project and guiding me through the process of performing my first research work. It is his friendly support and physical insights that made this voyage into nonlinear realms a most instructive and productive journey.

I would also like to thank Dr. Alex Retzker for all his help and the many hours of discussions. I thank Shmuel Marcovitch for his kind assistance and helpful comments, and Shai Machnes for the added value of QLib.

# Table of Contents

# Chapter 1.

# Overview

The ability to control quantum coherence in ion traps has increased dramatically in recent years. Following Feynman's visionary suggestion [1], quantum simulation schemes were proposed for the ion trap[2] [3] and realized [4]. In the context of Quantum Information Processing (QIP), the ion trap system is the arena of the well known Cirac-Zoller scheme [5]. High fidelity gates have been realized following this suggestion [6] and multi-partite entangled states have been created [7]. However, a straightforward scaling of these successes to a large number of ions seems to be extremely challenging [8].

In a different framework, nonlinear theories manifest classical non-perturbative configurations which are localized and topologically-stable, known as kinks. Linearizing the equations of motion around a static kink, a discrete set of localized eigenfunctions is found among the normal-modes. Perturbative quantization of these normal modes results in a spectrum of bound states.

The aforementioned characteristics of the discrete kink suggest that it may be advantageously utilized for QIP in a large system. The presence of the kink facilitates the cooling of a localized mode down to its ground state, as this process involves only a small number of participating particles. In addition to the vibrational degrees of freedom, the particles that comprise the system may possess internal levels. The spectral-isolation of the localized mode then enables to coherently couple this mode to the internal levels. Moreover, a translation of the kink may allow to transport quantum information around the system [9].

With the ion trap system in mind, we investigate the quantum-mechanical properties of fluctuations around topological excitations in discrete models. We treat the well-known sine-Gordon and $\phi^4$ models, in an approach which to our knowledge has not been previously employed. We analyze numerically the localized modes of a static kink solution focusing on the roles of discreteness and nonlinearity in a finite-size system. We use a non-Markovian master equation to solve the coherent time evolution of these modes coupled to a thermally-excited environment of linear-wave phonons. We discuss the possible application of our results to performing high-fidelity QIP operations in the presence of many ions.

The plan of this work goes as follows : In Chapter 2 we give a motivation for our work, shortly describing the fundamental present challenge in ion-trap QIP. In Chapter 3 we review basic properties of static solitons in quantum field-theory. These familiar results will echo throughout the next few sections. In Chapter 4 we start our research by examining the classical ground state of a kink in the two finite models studied therein. In Chapter 5 we exhibit the linearized analysis of such equilibrium configurations. We proceed with the nonlinear corrections in chapter Chapter 6, and start a dynamical analysis by solving the equations of motion in Chapter 7. Chapter 8 details the derivation of the master equation we use to solve the quantum-mechanical dynamics, and in Chapter 9 we present its numerical solutions. We conclude this work with a short discussion in Chapter 10.

# Chapter 2.

# Motivation – QIP in Ion Traps

In the common realizations of the Cirac-Zoller scheme for QIP in the ion trap [10] [11], singly-ionized alkaline-earth atoms are trapped by an effective harmonic potential[1]. The ground-state of each ion's single valence electron, together with a metastable level serve as a two-level system : a qubit for quantum computation. Operating on those qubits relies crucially on single addressing of each individual ion with a laser beam. In addition, two-qubit gates require the coherent manipulation of a vibrational phonon shared by different ions, used to couple their electronic levels. In the linear ion trap, the ions crystallize along the axis of least trapping frequency, and this phonon is chosen to be the center-of-mass (c.o.m) mode. It is lowest in the spectrum, has angular frequency 1 (in units of the trapping frequency along the axis) and is separated by a gap from the next phonon which has frequency $\sqrt{3}$.

The advantages of using the c.o.m phonon are its high decoupling from all the other modes in the trap and its ability to couple all ions in the trap. The disadvantages being its relatively high heating caused by external fields (which couple strongest to this mode) and, most importantly – the conflicting requirements from the frequency of the phonon (whose frequency is equal to the trapping frequency) ;

On the one hand, the trapping frequency scales inversely with (some power of) the number of trapped ions. Trapping more ions in a chain requires *lowering* the trapping frequency, in order to keep the density of ions above a certain limit. Otherwise, single addressing of ions would be impossible, and the chain might bifurcate into a two-dimensional zigzag configuration. On the other hand, the spectral isolation of the mode forms the basis of the ability to coherently access it with laser, independently of the other phonons. The gap separating this mode from the rest of the band (and also its purported gate-speed) are proportional to its frequency – the trapping frequency – and so is desired to be kept as *high* as possible. The current state-of-the-art for number in QIP experiments does not exceed eight ions [12]. The problem described "seems at the moment to be the biggest single problem standing in the way of advancing this field" [13].

---

[1] Trapping charged particles in three dimensions is impossible with only electrostatic fields, a result known as Earnshaw's theorem. In the linear Paul trap, ions are confined by rapidly oscillating quadrupole fields. When time-averaged, these give rise to an effective (secular) harmonic trapping.

Motivated by the above considerations, we investigate the coherence properties of a localized and gap-separated kink phonon. Being localized, its properties do not depend on the number of particles in the system. In particular, if realized in an ion trap, it should allow single addressing of ions without degradation of the gate-speed or the gap-separation as the number of ions in the trap is increased. This work is preliminary in that respect – we do not investigate the possibility of actually realizing a kink in the ion trap. We stay within the general framework of one-dimensional discrete kinks.

# Chapter 3.

# Review – Quantization of Static Kinks in Continuous Systems

We begin by discussing a few basic properties of nonlinear field-theory and the perturbative quantization of a static solitary wave solution. The central ideas presented in this section will form the basis of our later investigations, when we switch to a finite and discrete system. We follow the classical textbook on the subject by Rajaraman [14].

Consider then a single scalar field in 1+1 dimensions with the Lagrangian density

$$L = T[\phi] - V[\phi] = \frac{1}{2}(\phi_t)^2 - \left(\frac{1}{2}(\phi_x)^2 + U(\phi)\right). \tag{3.1}$$

The Euler-Lagrange equation of motion is the wave-equation

$$\phi_{tt} = \phi_{xx} - U_\phi. \tag{3.2}$$

We assume that $U \geq 0$ and that its absolute minima (finite or infinite, but enumerable), which are also its zeros, occur at

$$\theta_i \in \{U^{-1}(0)\}. \tag{3.3}$$

The total energy functional $E$ is given as an integral over the energy density, and is conserved over time;

$$E[\phi] = \int dx \left[\frac{1}{2}(\phi_t)^2 + \frac{1}{2}(\phi_x)^2 + U(\phi)\right]. \tag{3.4}$$

A detailed analysis shows that for a solution of the wave equation to be non-singular and have a finite energy and a localized energy density, it must obey the boundary conditions $\phi(+\infty) = \theta_i$, $\phi(-\infty) = \theta_j$. In addition, static (time-independent) solutions can only connect two adjacent minima of $U$. Finally, since $\phi(\pm\infty, t)$ is a continuous function of time, and has to equal one of the isolated zeros of $U$ at any time, it cannot 'jump' between values and is constrained to remain constant. Therefore the space of *all* non-singular finite-energy solutions divides naturally into sectors characterized by two indices – the time-independent values at the boundaries. These sectors are disconnected components of the function space, since no solution can be deformed continuously into another sector while remaining of finite-energy.

There is an important distinction between these conserved topological-indices and other conserved quantities like energy, momentum or charge. While the latter conservation laws are linked by Noether's theorem to continuous symmetries of the Lagrangian, the topological indices are simply boundary conditions, conserved when imposing finiteness of energy.

As two concrete examples, we take the double-well $\phi^4$ and the sine-Gordon (SG) equations,

$$\phi_{tt} = \phi_{xx} + \phi - \lambda\phi^3, \qquad (3.5)$$

$$\phi_{tt} = \phi_{xx} - \sin\phi. \qquad (3.6)$$

The $\phi^4$ potential has two degenerate minima values $\phi = \pm 1/\sqrt{\lambda}$, to which any solution must tend at infinity. Thus there are two topological sectors of solutions. We note that the equation admits the symmetry $\phi \to -\phi$ which switches the two minima.

The SG equation enjoys the same symmetry and also $\phi \to \phi + 2\pi i, \ i \in \mathbb{Z}$. This symmetry is the manifestation of the infinite number of minima of the sinusoidal potential $U(\phi) = 1 - \cos\phi$. There is an infinite number of topological sectors of solutions. The SG equation is actually quite exceptional among nonlinear PDEs, because of its *complete integrability* – any localized excitation can be represented as an asymptotic sum of elementary solutions of three types – phonons, solitons and breathers (which are not discussed here). This special property of the equation will reflect itself later in the numerical work we do with the corresponding discrete model.

It is about time we presented explicitly the one-soliton solutions of those equations, so that the label 'kink' used in the header becomes justified ;

$$\phi_{SG} = \pm \tan^{-1}\left[\exp(x - x_0)\right], \qquad (3.7)$$

$$\phi_{phi4} = \pm \frac{1}{\sqrt{\lambda}} \tanh\left[(x - x_0)/\sqrt{2}\right] \qquad (3.8)$$

These solutions are static kinks (or antikinks, for those with a minus sign in front), whose energy-density is centered around $x_0$. We show $\phi_{SG}$ in Figure 3.1.

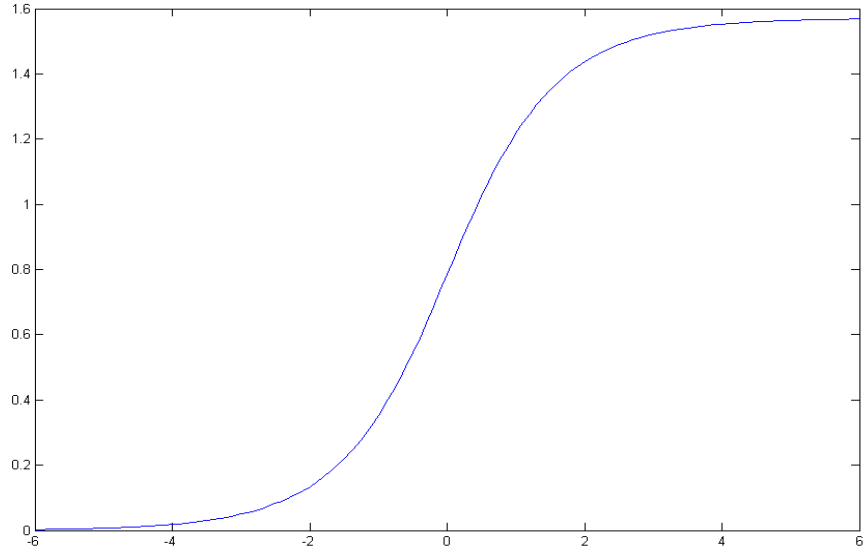

**Figure 3.1 : The sine-Gordon kink.**

A very important property of the kink is that it is singular as the nonlinearity parameter $\lambda$ tends to zero (in the SG solution this parameter is hidden by the choice of units, but it is there). Therefore it cannot be obtained in a perturbative expansion starting from the linearized version of the equations (a linear Kline-Gordon equation).

We turn now to the quantization of our scalar field. Let $\phi_0(x)$ be a stable static minimum of the potential $V[\phi] = \frac{1}{2}(\phi_x)^2 + U(\phi)$, satisfying $\delta V[\phi]/\delta\phi(x) = 0$. Expanding $V[\phi]$ in a functional Taylor expansion around $\phi_0(x)$ we get

$$V[\phi] = V[\phi_0] + \frac{1}{2}\int dx\, \eta(x)\left[-\partial_{xx} + \frac{d^2 U(\phi_0)}{d^2\phi}\right]\eta(x) + \ldots \tag{3.9}$$

where $\phi_0(x) \equiv \phi_0(x) + \eta(x)$. The quadratic form appearing in the integrand can be diagonalized using normal modes, the eigenfunctions of the Sturm-Liouville problem:

$$\left[-\partial_{xx} + \frac{d^2 U(\phi_0)}{d^2\phi}\right]\eta_i(x) = \omega_i^2 \eta_i(x). \tag{3.10}$$

In the resulting quantum field theory, the energy of any given state with occupation numbers $\{n_i\}$ will then be :

$$E\{n_i\} = V[\phi_0] + \sum_i \hbar\omega_i\left(n_i + \frac{1}{2}\right), \tag{3.11}$$

where $\hbar$ has been re-introduced. The higher-order corrections of interaction between the normal modes are to be treated in perturbation theory, and this is of course justified for small interaction coefficients.

Finally, we use the method outlined above to quantize the static kink (centered at $x_0 = 0$) of the $\phi^4$ theory, and get the eigenvalue equation

$$\left[-\frac{1}{2}\partial_{xx} + 3\tanh^2 x - 1\right]\eta_i(x) = \omega_i^2 \eta_i(x). \tag{3.12}$$

This Schrödinger-like equation has two bound levels followed by a continuum, from the bottom of which they were pulled-down ;

$$\omega_0^2 = 0, \qquad \eta_0(x) = 1/\cosh^2 x$$

$$\omega_1^2 = \frac{3}{2}, \qquad \eta_1(x) = \sinh x / \cosh^2 x$$

$$\omega_k^2 = 2 + \frac{1}{2}k^2, \qquad \eta_k(x) = e^{ikx}\left(3\tanh^2 x - 1 - k^2 - 3ik\tanh x\right) \tag{3.13}$$

The physical interpretation of these results follows.

1. The zero-mode is the result of the translation-invariance of the kink solution – we quantized around a specific "stable" minimum of the potential energy, while it is actually a family of solutions, which are only neutrally-stable. The zero-frequency eigenfunction is exactly the generator of infinitesimal translations of the kink.
2. The second discrete eigenfunction is spectrally separated by a gap from the continuum, and it is anti-symmetric around the kink center. It describes internal excitations of the kink – a shape-change mode.
3. The remaining states are just unit-mass plane-waves with some asymptotic phase-shift, and can be thought of as scattering states of the mesons of this theory, off the kink particle.

We note that the sine-Gordon model does not have in its spectrum an internal shape-change mode, but other than that it admits quite a similar analysis.

The properties of kinks and the process of their quantization as presented above for the case of quantum field theory, are fundamental and remain valid (with the required modifications) in the analysis we perform for the few-body problem.

# Chapter 4.

# The Static Discrete Kink

## *Introduction*

The Frenkel-Kontorova [15] model and its generalizations describe a chain of coupled particles subject to an external periodic on-site potential. An extensive body of analytical and numerical research of this model exists [16]. The chain can be finite or infinite, the boundary conditions periodic or fixed, the substrate potential simple (sinusoidal) or showing a complex pattern of multiple-wells and multiple-barriers. The particle interactions can be short-range or long-range, attractive or nonconvex (repulsive in part). These models admit different types of excitations – solitons (bright and dark), breathers, impurity modes and more. Unique phenomena and phases were discovered – the Aubry transition [17] and, in the quantum regime, the pinned instanton glass phase [18].

Regarding classical kink solutions, previous researches have focused mainly on characterizing the translational properties of the kink, on the zero-temperature behaviour or on the thermodynamic properties of a gas of quasi-excitations.

The effects of discreteness on the static kink are primarily the raising of the lowest-mode frequency to a finite non-zero value, and the existence of an upper cutoff frequency for the phonon band. The translational-mode frequency is the frequency of oscillation at the bottom of the Peierls-Nabarro potential, which captures the kink at a lattice-site. In some discrete models [19], one or more of the top frequency modes may also be 'lifted-up' from atop the cutoff frequency and become localized, gap-separated modes lying above the spectrum.

Another important effect enhanced by the discreteness is the radiation of phonons by moving kinks, by breathers and by kink internal modes. Such dissipative coupling is the major downside of the nonlinearity in these models, and will be at the heart of our investigation in this work.

## *The kink configuration*

In this part we will introduce the two models which this work investigates, and study shortly their classical stable configurations and their properties. The system we study is composed of a one-dimensional chain of particles, harmonically-coupled with their nearest-neighbors and subject also to an on-site nonlinear potential. We promote the coupling between neighbours to be location-dependent, which will allow us later a better control over the local Peierls-Nabarro trapping potential for the kink (as a result of the gradient of the coupling constants profile). We treat two different models : the sine-Gordon model, in which the particles can actually move between the sites (cross the barrier of the periodic potential), and the double-well $\phi^4$ model, where each oscillator is bound to a double-well potential from which it cannot escape.

We write all variables in non-dimensional units, with a minimum number of parameters entering the model. The physical units from which these derive depend on the underlying physical phenomena, and we will return to this point later as well. We start from the Lagrangian of *N* particles in standard form,

$$L = T - V = \sum_i^N \frac{1}{2} \dot{x}_i^2 - V \,. \tag{4.1}$$

We divide the potential energy into 3 parts : the coupling energy ($V_{coupling}$), the substrate potential ($V_{sub}$) and the boundary-conditions ($V_{bc}$).

For linearly-coupled oscillators we have

$$V_{coupling} = \sum_i^{N-1} \frac{g_i}{2} (x_{i+1} - x_i - a_0)^2 \tag{4.2}$$

where $g_i$ are location-dependent coupling constants and $a_0$ is the equilibrium length of the harmonic coupling spring.

For the double-well $\phi^4$ model :

$$V_{sub}^{(phi-4)} = \sum_i^N \left( \frac{1}{2} k x_i^2 + x_i^4 \right), \tag{4.3}$$

where *k* is negative. We use fixed-ends boundary-conditions to get one kink :

$$V_{bc}^{(phi-4)} = \frac{g_0}{2} (x_1 - a_1)^2 + \frac{g_0}{2} (x_N - a_N)^2 \,, \tag{4.4}$$

where $a_1 = \pm\sqrt{-k/4}$ and $a_N = -a_1$ are the zeros of the double-well and $g_0 \gg g_i$ binds the end-particles to their equilibrium locations.

For the sine-Gordon model,

$$V_{sub}^{(SG)} = \sum_{i}^{N} (1 - \cos x_i),$$ (4.5)

and we use periodic BC for a kink of topological charge $s=1$;

$$V_{bc}^{(SG)} = \frac{g_0}{2}\left(x_1 - x_N + (N+s-1)a_0\right)^2.$$ (4.6)

We find numerically the absolute minimum configuration of the potential energy. Subject to the aforementioned boundary conditions, the ground state of the chain is a one-kink configuration. In the figures below we show two configurations of kinks, one for each model.

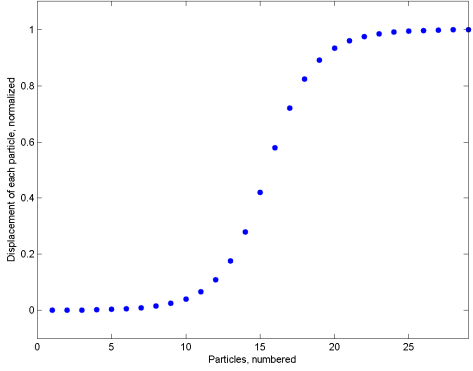

Figure 4.1 : A moderately-extended sine-Gordon kink, with $g_i \equiv G = 4$.

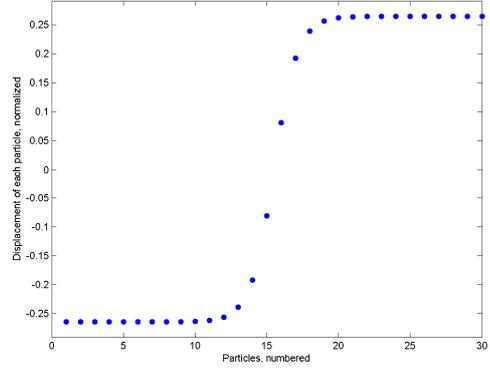

Figure 4.2 : A highly discrete kink of the $\phi^4$ model, with $g_i \equiv G = 0.4$, $k = -0.28$.

We now expand the potential energy around the equilibrium configuration locations of the particles $\{\bar{x}_0\}$;

$$V = \sum_{ij} \frac{1}{2!} K'_{ij} x_i x_j + \sum_{ijk} \frac{1}{3!} L'_{ijk} x_i x_j x_k + \sum_{ijkl} \frac{1}{4!} M'_{ijkl} x_i x_j x_k x_l + ...$$

$$K'_{ij} = \left.\frac{\partial^2 V}{\partial x_i \partial x_j}\right|_{\{\bar{x}_0\}} \quad L'_{ijk} = \left.\frac{\partial^3 V}{\partial x_i \partial x_j \partial x_k}\right|_{\{\bar{x}_0\}} \quad M'_{ijkl} = \left.\frac{\partial^4 V}{\partial x_i \partial x_j \partial x_k \partial x_l}\right|_{\{\bar{x}_0\}}$$ (4.7)

In the $\phi^4$ model there are no terms of higher order. In the sine-Gordon model the nonlinear terms have extremely good convergence properties (as we see later) and higher-order terms are completely irrelevant. With other forms of interaction between the particles (e.g. Coulomb interactions), the issue of convergence of these coefficients may be non-trivial.

The following chapters will be devoted to the analysis of the above expansion. Before continuing we note about the continuum limit of the models [20]. This limit corresponds to large $g_i$ (with the lattice constant $a_0$ absorbed), where the particles are strongly interacting and the difference of their displacements $(x_i - x_{i+1})/a_0$ is small. The kink in Figure 4.1 would then tend towards that of Figure 3.1.

# Chapter 5.

# Linear Analysis

The normal-modes of oscillation and their spectrum are found by diagonalizing the Hessian ($K'_{ij}$) of equation (4.7).

The normal-mode vectors are solutions of

$$\sum_j \left( K'_{ij} - \omega^2 \delta_{ij} \right) A_j = 0, \qquad \forall i. \tag{5.1}$$

This normalization corresponds to particles of unit mass.

The spectrum and the profile of some eigenvectors for the kink configurations which were shown in Figure 4.1 and Figure 4.2 are presented in the following figures. The end-modes are removed. We order the modes from the *highest* frequency mode *to the lowest*. This puts the localized modes, having the lowest frequencies, at the right end of the plot. The two localized modes do *not* have a long wavelength (and the translational mode does *not* obey $k \to 0$), and are not Goldstone modes. For the same reason we do not Fourier-transform the modes (plotting $\omega(k)$ etc.).

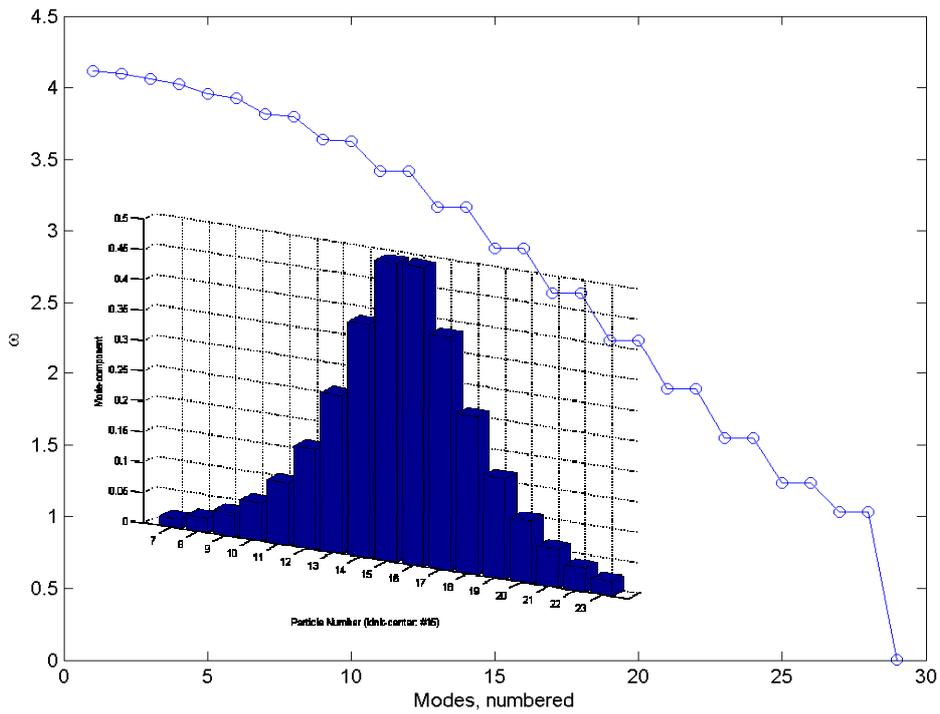

**Figure 5.1 : Dispersion relation for the SG kink, showing the frequencies of the modes ordered from highest to lowest. The degenerate couples are odd and even modes, result of the symmetry of the kink under reflection about its center. Inside : Spatial profile of the translational-mode.**

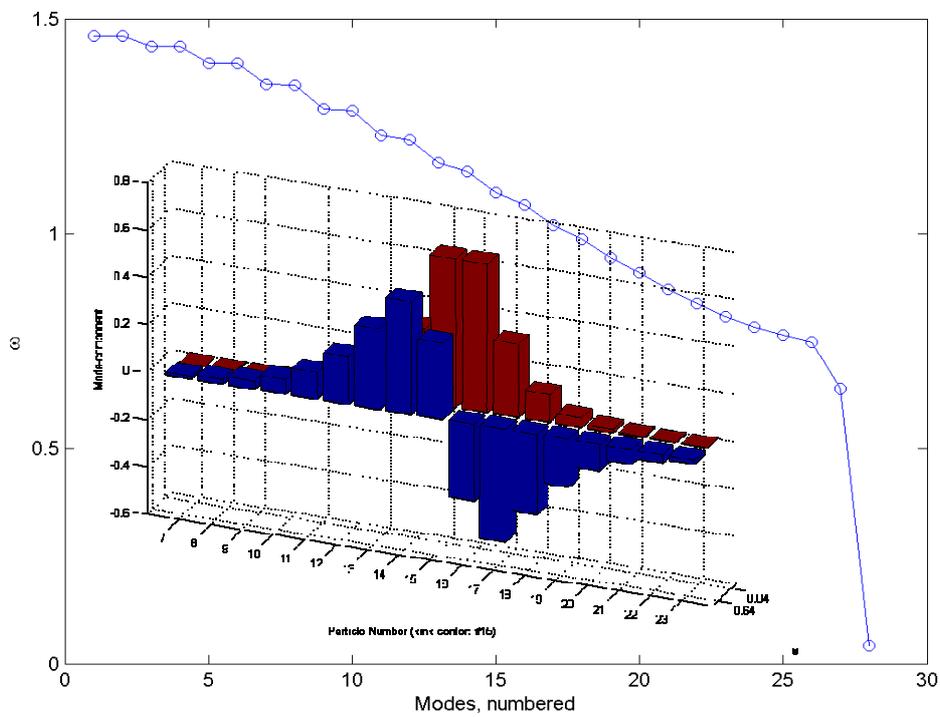

**Figure 5.2 : Dispersion relation for the $\phi^4$ kink. The gap separation is clearly seen. Inside : Profile of the two localized modes; the translational and the anti-symmetric.**

In Figure 5.3 we show the spectrum as obtained after assigning the particles' coupling coefficients with a location-dependent profile (shown in the inset). The kink excitation sees a very deep potential well centered at its core. This is manifested by lifting the spectrum up, in particular the translational-mode's frequency, $\omega_{low} \approx 0.276$.

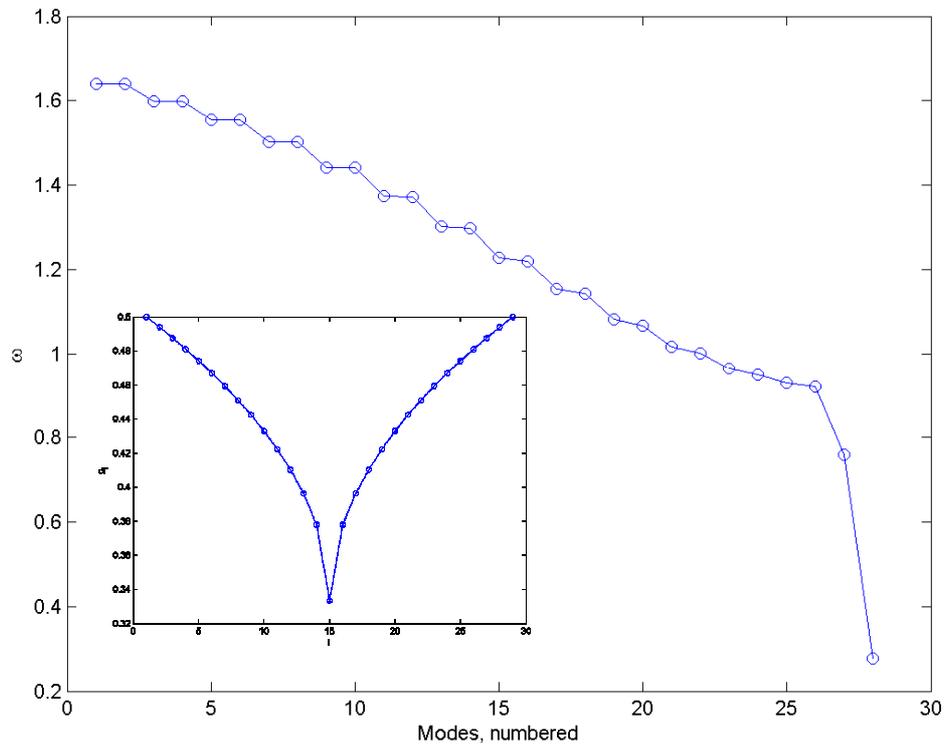

**Figure 5.3 : The spectrum of a kink with location-dependent coupling coefficients $g_i$.**

**Inside : $g_i$ as a function of the ion's number *i*.**

# Chapter 6.

# Nonlinear Couplings

If the orthonormal eigenvectors which diagonalize $K'_{ij}$ are given by the columns of $\lambda_{ij}$, then the potential energy in terms of normal coordinates $\Theta_j$ is

$$V = \frac{1}{2!}\sum_{ij} K_{ij}\Theta_i\Theta_j + \frac{1}{3!}\sum_{ijk} L_{ijk}\Theta_i\Theta_j\Theta_k + \frac{1}{4!}\sum_{ijkl} M_{ijkl}\Theta_i\Theta_j\Theta_k\Theta_l + ...$$

$$K_{ij} = \sum_{mn} K'_{mn}\lambda_{mi}\lambda_{nj} = \omega_i^2 \delta_{ij}$$

$$L_{ijk} = \sum_{mns} L'_{mns}\lambda_{mi}\lambda_{nj}\lambda_{sk}$$

$$M_{ijkl} = \sum_{mnst} M'_{mnst}\lambda_{mi}\lambda_{nj}\lambda_{sk}\lambda_{tl}$$

(6.1)

The first term is just a sum over non-interacting oscillators. The other terms are coupling terms describing multi-phonon processes. In the following figures we show examples of the matrices describing the coupling of the kink localized modes. The localized modes have the highest ordinal numbers (at the bottom-right corner).

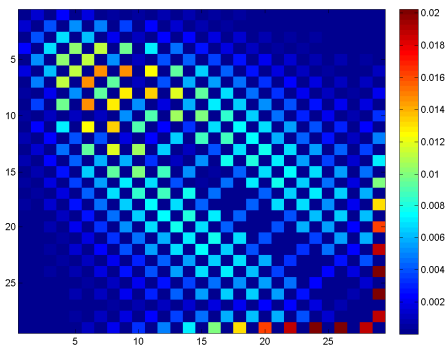 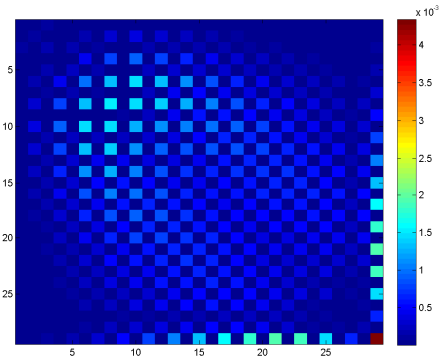

**Figure 6.1 : 3$^{rd}$ order coupling-strength of the SG kink *translational* mode. The matrix is $\left|L_{(*)jk}\right|/3!$, where * stands for this mode.**

**Figure 6.2 : 4$^{th}$ order coupling-strength of the SG kink *translational* mode. The matrix is $\left|M_{(**)jk}\right|/4!$, where * stands for this mode.**

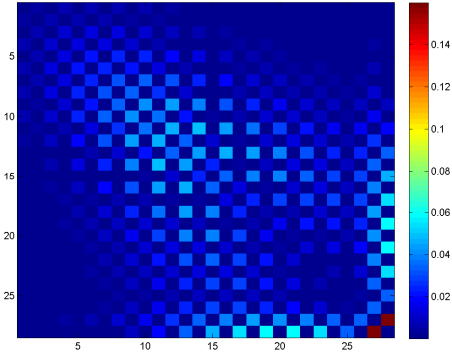
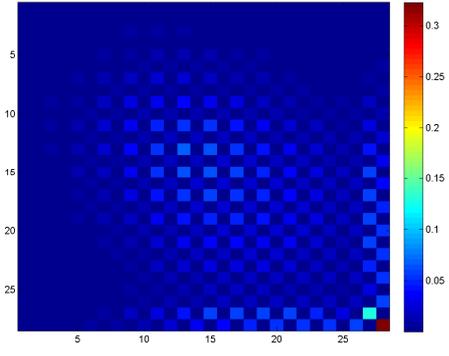

**Figure 6.3 :** 3$^{rd}$ order coupling-strength of the $\phi^4$ kink *translational* mode. The matrix is $|L_{(*)jk}|/3!$, where * stands for this mode.

**Figure 6.4 :** 4$^{th}$ order coupling-strength of the $\phi^4$ kink *translational* mode. The matrix is $|M_{(**)jk}|/4!$, where * stands for this mode.

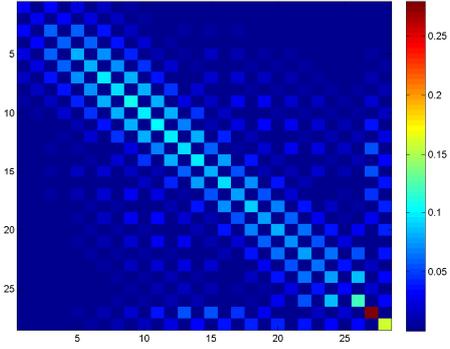
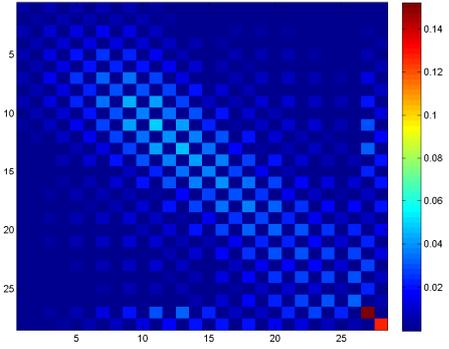

**Figure 6.5 :** 3$^{rd}$ order coupling-strength of the $\phi^4$ kink *anti-symmetric* mode. The matrix is $|L_{(*)jk}|/3!$, where * stands for this mode.

**Figure 6.6 :** 4$^{th}$ order coupling-strength of the $\phi^4$ kink *anti-symmetric* mode. The matrix is $|M_{(**)jk}|/4!$, where * stands for this mode.

It is immediately seen that the nonlinear terms are strong for the local modes. Both the self-interaction and the coupling coefficients are significant, making the oscillators really anharmonic. The chessboard structure follows from the modes being parity eigenstates, as the Hamiltonian is parity-invariant. It is also clearly visible that the coupling terms in the $\phi^4$ model are much larger than those in the sine-Gordon model. In the latter model there is also a decrease by an order of magnitude in the coefficients of 4$^{th}$ order compared to the 3$^{rd}$ order. We attribute this property to the complete-integrability of the model in the continuum-limit.

The next chapter will be devoted to a classical exploration of the strength of this nonlinearity.

# Chapter 7.

# Time Evolution

## *Basic goal and scheme*

As a first step towards understanding the dynamics governing the time evolution of the modes, we turn to a numerical simulation of the equations of motion, using simple and computationally-cheap tools (as compared with simulation of quantum-mechanical dynamics). The analysis we perform in this section allows us a first peek at the regime in phase-space where we are operating. We reach some important insights as to the nonlinear phenomena and the pitfalls awaiting.

In the sine-Gordon model the equations of motion read

$$\ddot{x}_i = -\left[ -g_i(x_{i+1} - x_i - a_0) + g_{i-1}(x_i - x_{i-1} - a_0) + \sin x_i + g_0(x_1 - x_N + (N+s-1)a_0)(\delta_{i1} - \delta_{iN}) \right], \quad (7.1)$$

while in the $\phi^4$ model they are

$$\ddot{x}_i = -\left[ -g_i(x_{i+1} - x_i - a_0) + g_{i-1}(x_i - x_{i-1} - a_0) + kx_i + 4x_i^3 + (x_1 - a_1)\delta_{i1} + g_0(x_N - a_N)\delta_{iN} \right]. \quad (7.2)$$

(Terms with indices *0* and *N+1* are to be dropped.)

The physical coordinates $x_i$ and the normal coordinates $\Theta_j$ are related by

$$x_i(t) = \sum_j \lambda_{ij}\Theta_j(t), \quad \Theta_j(t) = \Theta_j(0)e^{i\omega_j t} \quad (7.3)$$

(where $\lambda_{ij}$ are the normal-mode vectors).

Inverting the coordinate transformation and differentiating with respect to time *t*

$$\Theta_j(t) = \sum_i \lambda_{ij} x_i(t), \quad \dot{\Theta}_j(t) = \sum_i \lambda_{ij} \dot{x}_i(t). \quad (7.4)$$

The energy of a mode at any time is given by

$$E_j(t) = \frac{1}{2}\left(\dot{\Theta}_j^2 + \omega_j^2 \Theta_j^2\right) = \frac{1}{2}\sum_{ik} \lambda_{ij}\lambda_{kj}\left(\omega_j^2 x_i x_k + \dot{x}_i \dot{x}_k\right). \quad (7.5)$$

To develop our intuition for the behaviour of the nonlinear oscillators at the regime of energies which are of interest to us, we first turn to use a semi-classical language. We will simulate the *classical* dynamics of the system as outlined above, but assign initial conditions and analyze the energies by using quantum *phononic* terminology.

Thus we here introduce Planck's constant $\hbar$ in the nondimensional units of our models. We obviously cannot pull a numeric value out of the hat, but since this work is motivated by a specific physical system (namely the ion trap) we can get a natural value from this system, and we use $\hbar \approx 1.9 \cdot 10^{-5}$. The exact value depends on the ions used and the inter-ion distance. The value taken here corresponds to $Ca^+$ ions with an inter-ion separation $\sim 2\mu$, at trapping frequency of $20 MHz$. With this value of $\hbar$, the particles wave-packets have a width of just a fraction of the lattice-distance[2]. This justifies treating the particles as classical and distinguishable, and the chain as having a unique ground state, with tunneling effects being negligible.

The number of phonons in a mode is extracted from the energy in a mode by dividing the latter by $\hbar\omega_j$. In order to raise *n* phonons of a mode *j* (no zero-motion considered here), we use the semi-classical initial conditions

$$\frac{1}{2}\dot{\vec{x}}^2 = E_{kinetic} = n\hbar\omega_j, \qquad (7.6)$$

where $\dot{\vec{x}}^2$ is the velocity vector of *all the particles*. This translates to particle coordinates as

$$\dot{x}_i = \lambda_{ij}\sqrt{2E_j} = \lambda_{ij}\sqrt{2n\hbar\omega_j}. \qquad (7.7)$$

We will also consider the dynamics when all the modes are subject to a thermal distribution of energy. In order to populate the modes at a temperature *T* we use the Bose-Einstein distribution and measure the temperature in units of $\hbar$. In the ion-trap, the Doppler cooling[3] limit corresponds in our units to $T \sim 0.5$. The distribution is simply expressed as

$$\langle n_j \rangle = \left(e^{\omega_j/T} - 1\right)^{-1}. \qquad (7.8)$$

---

[2] For the harmonic oscillator : $\langle x^2 \rangle = \hbar(n+1/2)/m\omega$, and we treat temperatures which admit a few phonons at the most.

[3] Doppler laser cooling is based on the vibrating ions scattering photons from a red-detuned laser beam. The Doppler effect makes the ions 'see' the laser wavelength as on-resonance only in the part of their motion when the laser is counter-propagating theirs. The scattering is then enhanced anisotropically and the ions cool down to (approximately) the line-width of the transition.

Choosing this distribution is convenient since we are interested in the short-time behaviour of the system. We are not investigating the thermodynamics of the system, thus we do not go into complicated considerations of the system's classical long-term behaviour, such as equipartition of energy. We are interested in a classical simulation at the regime of energies relevant to the quantum-mechanical system.

One last tool we will employ is Fourier analysis. For a single coordinate sampled at discrete equidistant points in time : $x(j) \equiv x(t_j), \quad j = 1,...,L$, over a total time duration $\tau$, the discrete Fourier transform is defined by the vector

$$F(k) = \sum_{j=1}^{L} x(j) e^{-2\pi i (k-1)(j-1)/L} .$$

(7.9)

Entries of this vector are related to the frequencies by

$$\omega_k = 2\pi(k-1)/\tau, \qquad k = 2,...,L/2 .$$

(7.10)

($F(1)$ is just the sum of the input data, $F(L/2)$ is the Nyquist frequency, and above that the coefficients are for negative frequencies, i.e. irrelevant for positive input).

## *Building nonlinear intuition*

We start by simulating the dynamics in the $\phi^4$ model in the configuration of Figure 4.2 and Figure 5.2. The e.o.m of all the particles are solved as a function of time, from which we get the solutions for the modes. In Figure 7.1 and Figure 7.2 we show results for the simplest initial condition – just one phonon of the anti-symmetric mode (the high mode) is excited. The behaviour of the mode is harmonic.

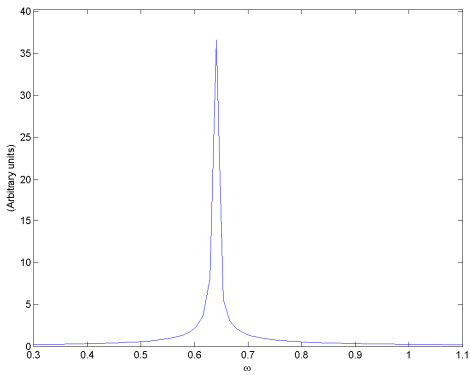

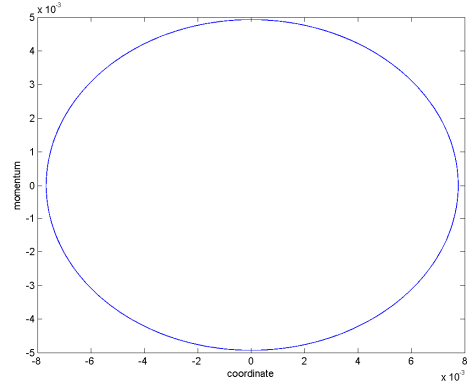

**Figure 7.1 : Discrete Fourier Transform of the high mode coordinate, one excited phonon. A single clear peak is centered around $\omega \approx 0.64$.**

**Figure 7.2 : Phase space trajectory of the high mode, one excited phonon. It is clearly a harmonic-oscillator's ellipse.**

We expect that in order to see a manifestation of the mode's self-nonlinearity, we must increase its energy. It turns out that we need to go to much higher energies to get clear nonlinear phenomena of this mode alone. In the following figures we show its behaviour with 200 phonons excited (but no other mode excited).

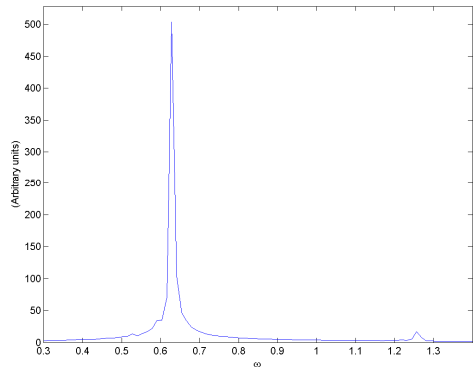

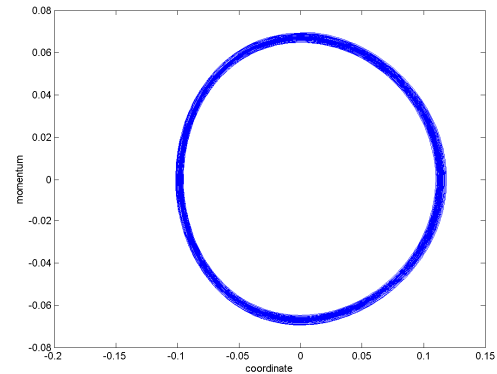

**Figure 7.3 : Discrete Fourier transform of the high mode coordinate, 200 excited phonons. The major peak is centered around** $\omega \approx 0.628$. **A minor peak appears at the second harmonic.**

**Figure 7.4 : Phase space trajectory of the high mode, 200 excited phonons. It is clearly an anharmonic oscillator now.**

We now return to low energies and assign initial conditions of one phonon of the anti-symmetric mode (the high mode) and four phonons of the translational mode (the low mode). Since the ratio of their frequencies is $0.64/0.043 \approx 15$, the low mode is excited with only $\sim 1/4$ of the energy of the high one. Energy in the modes is simulated as a function of time. In the following plots we divide time by $2\pi$ so that one unit on the time axis corresponds to one period of oscillation of frequency $1$.

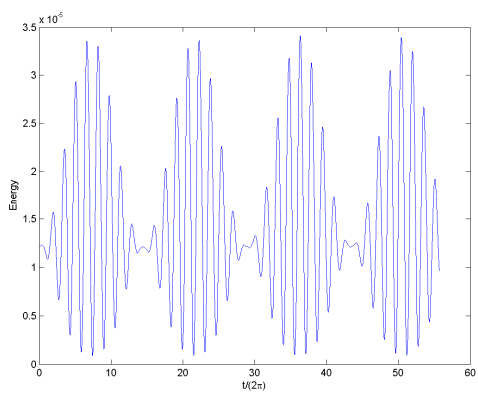

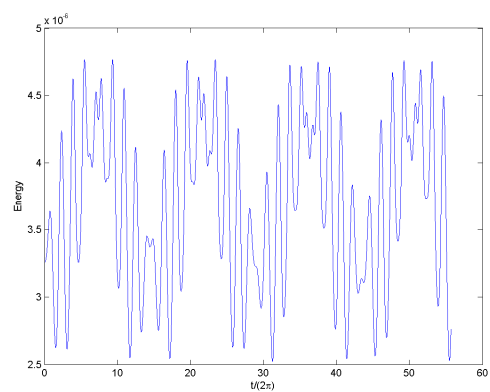

**Figure 7.5 : Energy in the high mode, as function of time.**

**Figure 7.6 : Energy in the translational (low) mode, as function of time.**

It is immediately clear that something is misbehaved.. Were the modes truly harmonic, we would expect a constant line in both figures. But the energy in both modes performs huge oscillations, with both frequencies present! Presumably, the energy is not even conserved. However, the total energy of the particles is constant up to the numerical accuracy $\sim 10^{-13}$. In Figure 7.7 we show the energy only in the non-localized modes. We see that it is almost the same as the low mode's energy, and oscillates with its frequency. In Figure 7.8 we show the total (harmonic) energy in all modes. This graph is dominated by the contribution from the high mode (compare Figure 7.5), with a small shift in energy caused by the sum of all other modes (sum of Figure 7.6 and Figure 7.7).

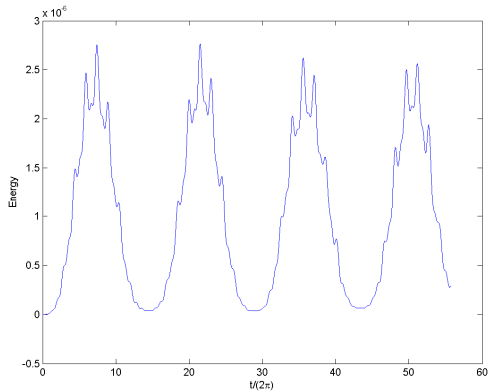 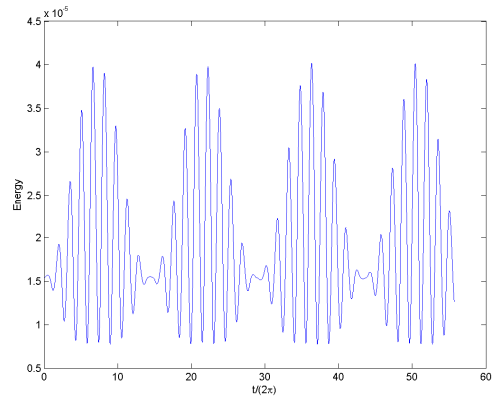

**Figure 7.7 : Energy in all non-localized modes.**    **Figure 7.8 : Energy in all modes.**

So we discover that a lot of energy hides inside the anharmonic terms, especially in the couplings of the low mode, and mostly in its coupling with the high mode. We now add to the above initial conditions, a thermal (Bose-Einstein) excitation of all the other modes, at temperature $T=1$ (in units of $\hbar$ as discussed above). We show the excitation used as initial conditions for this run. We then show the energy of the low mode and the phase-space trajectories of the localized modes.

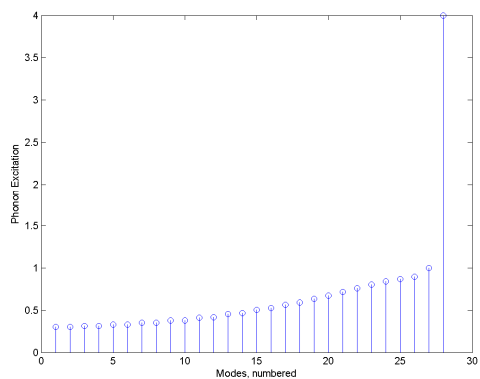 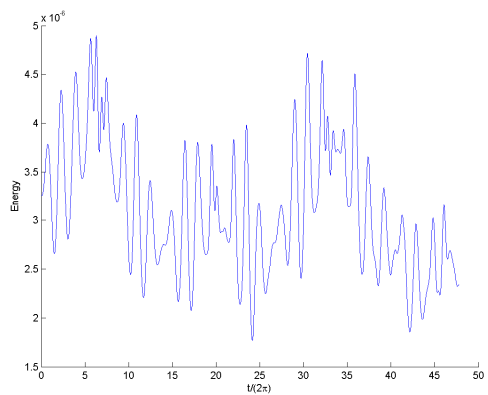

**Figure 7.9 : Initial conditions of excitation in the modes, at $T=1$. The two localized modes are populated with 4 and 1 phonons.**    **Figure 7.10 : Energy in the low mode, as function of time. Interaction with the thermal environment shows an erratic pattern.**

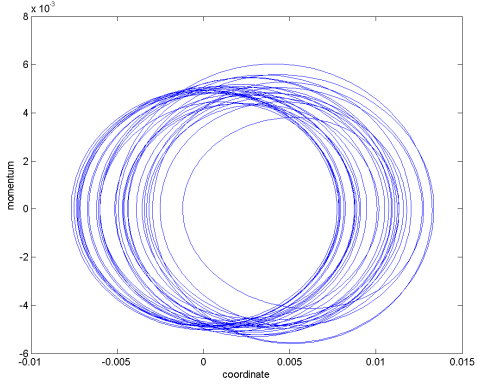
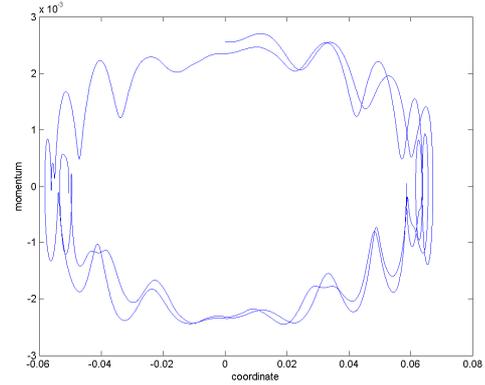

**Figure 7.11 : Phase space trajectory of the high mode, for the initial conditions of Figure 7.9.**

**Figure 7.12 : Phase space trajectory of the low mode, for the initial conditions of Figure 7.9.**

We now examine the configuration introduced in Figure 5.3, with $\omega_{low} \approx 0.276$ and $\omega_{high} \approx 0.758$. We introduce a thermal excitation into the modes including the low mode at temperature $T = 0.5$ (but the high mode is excited with *one* phonon). We show the excitation of the modes used as initial conditions for the simulation. In the following figures it is seen that the high mode's energy oscillates about its mean value, with roughly ±15% of the energy going 'in and out' of the harmonic terms, and the average stays almost constant – almost no energy is dissipated.

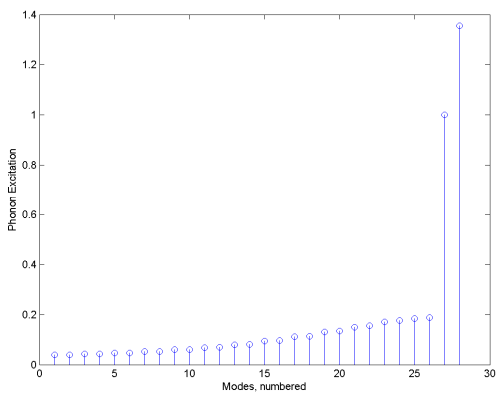
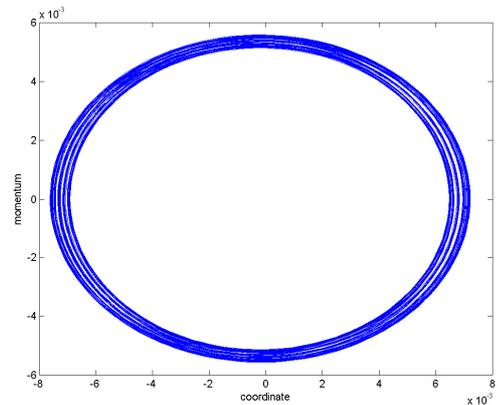

**Figure 7.13 : Thermal (Bose-Einstein) excitation of the modes used as the initial condition. The high localized mode (second-to-last) is excited with one phonon.**

**Figure 7.14 : Phase space trajectory of the high mode for the simulation. Although not a thin harmonic-ellipse, it is still rather acceptable as an oscillator with small corrections.**

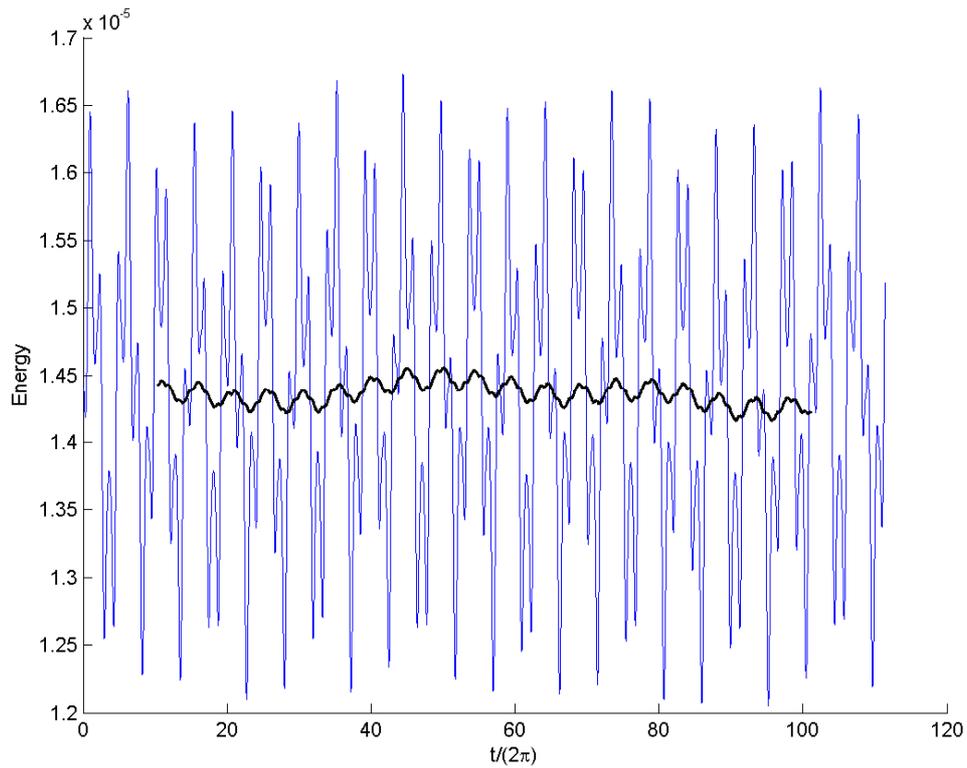

**Figure 7.15 : Energy of the high mode and its running average. It shows periodic fast oscillations of a fraction of the energy. Almost no energy seems to dissipate away (or be absorbed).**

We conclude by running a similar simulation of e.o.m in the sine-Gordon model. Even with $T=1$, the single localized mode shows a very high degree of linearity and isolation.

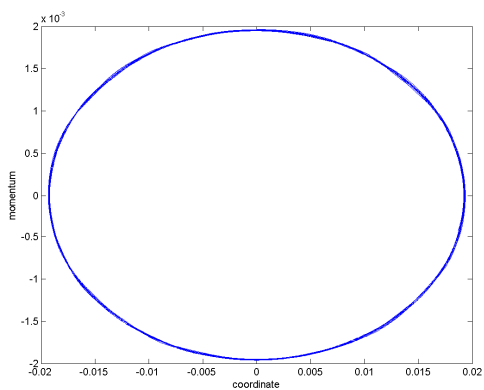
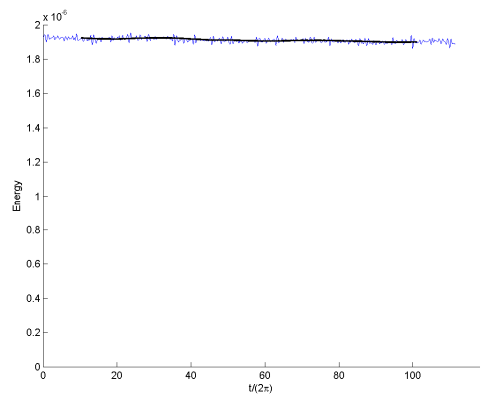

**Figure 7.16 : Phase space trajectory of one sine-Gordon localized phonon, at the presence of thermal environment with $T=1$.**

**Figure 7.17 : Energy of one sine-Gordon localized phonon as function of time, at the presence of thermal environment with $T=1$.**

# Chapter 8.

# Quantization and Master Equation

We quantize our Hamiltonian by the canonical quantization procedure for oscillatory normal-modes

$$\Theta_i \to \sqrt{\hbar/2\omega_i}\left(a_i^\dagger + a_i\right), \qquad \dot{\Theta}_i \to i\sqrt{\hbar\omega_i/2}\left(a_i^\dagger - a_i\right), \qquad (8.1)$$

where $a_i^\dagger$ are creation operators for the mode $i$.

We get

$$H = \sum_i \hbar\omega_i\left(a_i^\dagger a_i + \frac{1}{2}\right) + \frac{1}{3!}\sum_{ijk} L_{ijk}\hbar^{3/2}\left(8\omega_i\omega_j\omega_k\right)^{-1/2}\left(a_i^\dagger + a_i\right)\left(a_j^\dagger + a_j\right)\left(a_k^\dagger + a_k\right) +$$
$$+ \frac{1}{4!}\sum_{ijkl} M_{ijkl}\hbar^2\left(16\omega_i\omega_j\omega_k\omega_l\right)^{-1/2}\left(a_i^\dagger + a_i\right)\left(a_j^\dagger + a_j\right)\left(a_k^\dagger + a_k\right)\left(a_l^\dagger + a_l\right) + \ldots \qquad (8.2)$$

We truncate the Hamiltonian at fourth order as explained in the previous sections. This truncation will be further discussed at the last section.

We now write a master equation to model the coherent quantum-mechanical evolution in time of the phonon of interest. We divide the modes into the 'system' – consisting of the two coupled nonlinear oscillators, which we re-number as 1 and 2, and the 'bath' – all other modes. We split the Hamiltonian $H = H_S + H_B + H_{SB}$ into 3 parts ; for the system, the bath and the system-bath interaction. The Liouville-von Neumann equation for the combined system and bath density-operator $\Upsilon(t)$ reads

$$\dot{\Upsilon}(t) = -\frac{1}{\hbar}\left[H, \Upsilon(t)\right]. \qquad (8.3)$$

Going to the interaction-picture by exponentiating $H_S + H_B$ we define

$$\tilde{\Upsilon}(t) \equiv e^{i(H_S + H_B)t/\hbar}\Upsilon(t)e^{-i(H_S + H_B)t/\hbar}. \qquad (8.4)$$

A few manipulations results in the following (still-exact) equation

$$\dot{\tilde{\Upsilon}}(t) = \frac{1}{i\hbar}\left[\tilde{H}_{SB}(t), \Upsilon(0)\right] - \frac{1}{\hbar^2}\left[\tilde{H}_{SB}(t), \int_0^t dt'\left[\tilde{H}_{SB}(t'), \tilde{\Upsilon}(t')\right]\right]. \qquad (8.5)$$

We will assume that at time $t = 0$ there is no entanglement between the system and the bath, i.e. $\Upsilon(0)$ factorizes as

$$\Upsilon(0) = \rho(0) \otimes \chi_B, \qquad (8.6)$$

where $\rho(t)$ is the system's density-operator and $\chi_B$ is the initial density-operator of the bath. We now make the Born approximation to get an equation for the system's density matrix $\rho(t)$ alone. This means that the interaction Hamiltonian appears to second order in an integral expansion of $\dot{\rho}(t)$, or equivalently – that there is no evolution of the bath's density matrix [21]. We can thus substitute $\tilde{\Upsilon}(t) = \tilde{\rho}(t) \otimes \chi_B + O(H_{SB})$, and trace over the bath to get the master equation

$$\dot{\tilde{\rho}}(t) = -\frac{1}{\hbar^2} tr_B \left\{ \left[ \tilde{H}_{SB}(t), \int_0^t dt' \left[ \tilde{H}_{SB}(t'), \tilde{\rho}(t') \chi_B \right] \right] \right\}. \qquad (8.7)$$

We here eliminated the term linear in $\tilde{H}_{SB}(t)$, by assuming

$$tr_B \left\{ \tilde{H}_{SB}(t) \chi_B \right\} = 0. \qquad (8.8)$$

We will return to this point shortly. We write the system-bath interaction as a sum of terms [22], each containing a system operator ($s_\alpha$) multiplying a bath operator ($B_\alpha$):

$$\tilde{H}_{SB}(t) = \hbar \sum_\alpha \tilde{s}_\alpha(t) \tilde{B}_\alpha(t) = \hbar \sum_\alpha e^{iH_S t/\hbar} s_\alpha e^{-iH_S t/\hbar} e^{iH_B t/\hbar} B_\alpha e^{-iH_B t/\hbar}. \qquad (8.9)$$

With this definition (using the cyclic property of the trace)

$$\dot{\tilde{\rho}}(t) = -\sum_{\alpha,\beta} \int_0^t dt' \left\{ \begin{array}{l} \left[ \tilde{s}_\alpha(t) \tilde{s}_\beta(t') \tilde{\rho}(t') - \tilde{s}_\beta(t') \tilde{\rho}(t') \tilde{s}_\alpha(t) \right] \langle \tilde{B}_\alpha(t) \tilde{B}_\beta(t') \rangle_B + \\ \left[ \tilde{\rho}(t') \tilde{s}_\beta(t') \tilde{s}_\alpha(t) - \tilde{s}_\alpha(t) \tilde{\rho}(t') \tilde{s}_\beta(t') \right] \langle \tilde{B}_\beta(t') \tilde{B}_\alpha(t) \rangle_B \end{array} \right\}, \qquad (8.10)$$

where we defined the bath correlation functions

$$\langle \tilde{B}_\alpha(t) \tilde{B}_\beta(t') \rangle_B = tr_B \left\{ \chi_B \tilde{B}_\alpha(t) \tilde{B}_\beta(t') \right\}. \qquad (8.11)$$

In the bath Hamiltonian there is no need to take nonlinear coupling terms, as we assume the bath to stay in thermal equilibrium. In the interaction terms, we take bath operators no higher than quadratic, interacting with linear and quadratic system operators

$$s_\alpha \in \left\{ (a_1^\dagger + a_1), (a_2^\dagger + a_2), (a_1^\dagger + a_1)^2, (a_2^\dagger + a_2)^2, (a_1^\dagger + a_1)(a_2^\dagger + a_2) \right\}. \qquad (8.12)$$

We here neglected three possible interactions coming from the fourth-order term in the Hamiltonian, containing three system operators interacting with one bath operator, the largest of these being $\left(a_2^\dagger + a_2\right)^3 \left(a_j^\dagger + a_j\right)$. But examining the coupling coefficients shows that the latter are not of exceptional magnitude, and in any case there are only $\sim N$ of them, compared to the $\sim N^2$ other interactions. The other two terms of the form $\left(a_1^\dagger + a_1\right)\left(a_2^\dagger + a_2\right)^2 \left(a_j^\dagger + a_j\right)$ are diminished in coefficient but might (a-priori) show an excpetional resonance condition. This can be avoided (as discussed in the conclusion).

There are 5 corresponding bath operators

$$\tilde{B}_i(t) = 3 \cdot \frac{1}{3!} \sum_{\substack{i \in \{1,2\} \\ jk \neq 1,2}} L_{ijk} \hbar^{1/2} \left(8\omega_i \omega_j \omega_k\right)^{-1/2} \left(\tilde{a}_j^\dagger + \tilde{a}_j\right)\left(\tilde{a}_k^\dagger + \tilde{a}_k\right). \quad (8.13)$$

$$\tilde{B}_{ij}(t) = 3 \cdot \frac{1}{3!} \sum_{\substack{i,j \in \{1,2\} \\ k \neq 1,2}} L_{ijk} \hbar^{1/2} \left(8\omega_i \omega_j \omega_k\right)^{-1/2} \left(\tilde{a}_k^\dagger + \tilde{a}_k\right) + 6 \cdot \frac{1}{4!} \sum_{kl \neq 1,2} M_{ijkl} \hbar \left(16\omega_i \omega_j \omega_k \omega_l\right)^{-1/2} \ldots \quad (8.14)$$

($\tilde{B}_{12}(t)$ is not a different operator from $\tilde{B}_{21}(t)$, it is just an ambiguity of the notation. We have indeed five operators).

The bath operators obey $\tilde{a}_k(t) = a_k e^{-i\omega_k t}, \tilde{a}_k^\dagger(t) = a_k^\dagger e^{i\omega_k t}$, so with $\chi_B$ a thermal density-operator, which commutes with the bath Hamiltonian, the only survivors of bath expectation values are

$$tr_B\left\{\chi_B a_j^\dagger a_k\right\} = \delta_{jk} n_j, \qquad tr_B\left\{\chi_B a_j a_k^\dagger\right\} = \delta_{jk}(n_j + 1). \quad (8.15)$$

($n_j \equiv n_j(\omega_j, T)$ being the mean excitation of bath-mode $j$).

In appendix A we write the correlation functions explicitly, and show how the system Hamiltonian is renormalized to satisfy condition (8.8) above. The renormalized correlation functions are redefined as $\tilde{C}_{\alpha\beta}(t-t')$, and we use the latter in the master equation from this point on. The master equation is now

$$\dot{\tilde{\rho}}(t) = -\sum_{\alpha,\beta} \int_0^t dt' \left\{ \begin{array}{l} \left[\tilde{s}_\alpha(t)\tilde{s}_\beta(t')\tilde{\rho}(t') - \tilde{s}_\beta(t')\tilde{\rho}(t')\tilde{s}_\alpha(t)\right]\tilde{C}_{\alpha\beta}(t-t') + \\ \left[\tilde{\rho}(t')\tilde{s}_\beta(t')\tilde{s}_\alpha(t) - \tilde{s}_\alpha(t)\tilde{\rho}(t')\tilde{s}_\beta(t')\right]\tilde{C}_{\beta\alpha}(t'-t) \end{array} \right\}. \quad (8.16)$$

# Chapter 9.

# Numerics with the Master Equation

We can now solve the master equation (8.16) by numerical integration. This is done in the interaction picture, with the full Hamiltonian of the system propagating its operators. We then have the system's density matrix in the interaction picture, which can be taken back into Schrödinger picture. In this picture we trace-over the low mode's degree of freedom to get the reduced density matrix of the high mode alone.

However, the time required to integrate an integro-differential equation scales like $\sim n^2$ with $n$ being the number of time-steps required in an ordinary differential equation, and this may be quite demanding on computer resources. With the system having five operators, the operator-sum in equation (8.16) contains $4 \cdot 5^2 = 100$ terms. Each term is a multiplication of three matrices. The size of the matrices depends on the dimension at which the Hilbert-space of each of the system's oscillators is truncated, and it is at least *7*, making each matrix size at least $7^2 \times 7^2$ (It is verified that the truncation of the Hilbert space has a negligible impact on the accuracy of our calculation).

For the simulation we choose a configuration which is quite similar to that presented in Figure 6.3. We increase the number of particles to *90*. Recall that we abandoned the numbering of the modes as used throughout the first part of this work. The anti-symmetric (high) mode is re-numbered as mode number *1*, and the translational (low) mode as number *2*. The system in the master equation can therefore be composed of mode number *1* alone (the low mode is then included in the bath), or of both modes – *1* and *2*. The frequencies are $\omega_1 \approx 0.743$ and $\omega_2 \approx 0.274$. We next show the most significant bath correlation function, for both cases – with the low mode included either in the bath or in the system :

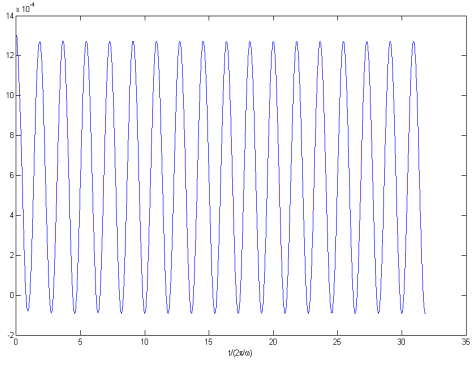
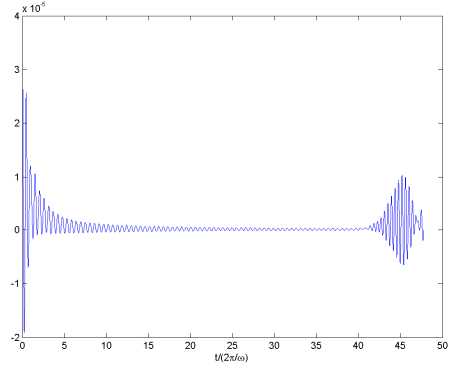

**Figure 9.1 : Real part of the renormalized bath correlation function $\tilde{C}_{11}(\tau)$, with the low mode ($\omega_2$) included in the bath.**

**Figure 9.2 : Real part of the renormalized bath correlation function $\tilde{C}_{11}(\tau)$, with the low mode ($\omega_2$) included in the system.**

We continue in three paths of investigation.

1. Solve the dynamics as presented above, with the system composed of the two localized modes, without assuming Markovicity of the bath, for a simulation-time as long as computer resources allow.

2. A second option we follow is to take the low mode out of the system and put it into the bath. This leaves only two system operators (linear and quadratic in the one mode of interest), each being a matrix of a much smaller size (the Hilbert space now contains only one oscillator). We then solve the numerics as presented above, without any further approximation.

3. A third option, motivated by Figure 9.2 above, is to assume that the bath correlation functions vanish after (e.g.) $\tau = 15$, and so perform the integration for this time-span only. We neglect the rise in correlation shown at times starting with $\tau \approx 40$ (and again at longer times, not shown). This option is obviously relevant only for the system with both modes included.

We perform the time integrations formally over the system's history, using the symmetry property of $\tilde{C}_{\alpha\beta}(t-t')$ under exchange of the indices and the relation $\tilde{C}_{\alpha\beta}(-\tau) = \tilde{C}_{\alpha\beta}^{*}(\tau)$. Using the definition

$$\tilde{S}_{\alpha\beta}(t) \equiv \int_0^t \tilde{s}_\alpha(t') \tilde{\rho}(t') \tilde{C}_{\alpha\beta}(t-t') dt' = \int_0^t e^{iH_s t'/\hbar} s_\alpha e^{-iH_s t'/\hbar} \tilde{\rho}(t') \tilde{C}_{\alpha\beta}(t-t') dt',$$
(9.1)

We get the misleadingly simple equation

$$\dot{\tilde{\rho}} = -\sum_{\alpha,\beta} \left[ \tilde{s}_\alpha \tilde{S}_{\beta\alpha} - \tilde{S}_{\beta\alpha} \tilde{s}_\alpha + \tilde{S}_{\beta\alpha}^\dagger \tilde{s}_\alpha - \tilde{s}_\alpha \tilde{S}_{\beta\alpha}^\dagger \right].$$
(9.2)

We now show the results of simulations using the approximations discussed above. We use as initial conditions a superposition of the ground- and first excited-states in the high mode ($|0\rangle+|1\rangle$), two phonons in the low mode and a Bose-Einstein distribution (similar to that in Figure 7.13) in the other modes. Were it not for the nonlinear terms, the high mode would perform Rabi oscillations between its first and excited levels. We check the fidelity[4] of the high-mode's simulated density matrix compared with the density matrix of an isolated phonon subject only to the free Hamiltonian of the mode and performing such Rabi oscillations.

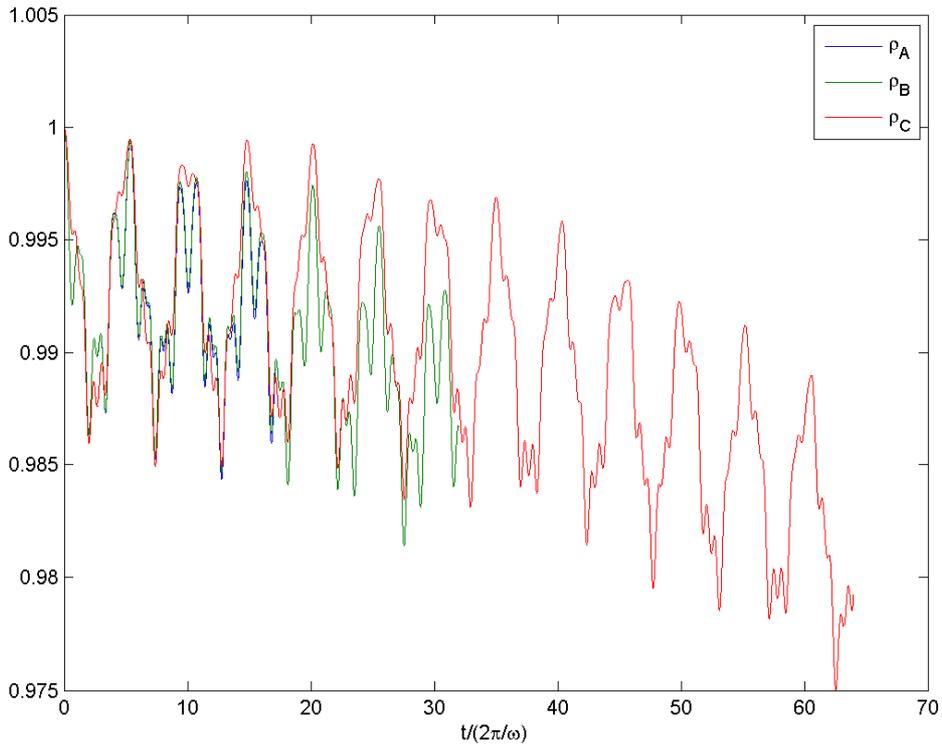

**Figure 9.3 : The fidelity of the high-mode phonon's Rabi oscillations vs. time, as a result of three simulations run. $\rho_A$ is the result of a full non-Markovian solution with both localized modes included in the system Hamiltonian. $\rho_B$ is the result of a full solution with both modes in the system, however with the correlation functions truncated after $\tau = 15$. $\rho_C$ is the result of a non-Markovian solution with the low mode included in the bath rather than the system.**

---

[4] The fidelity distance-measure between two density operators $\rho$ and $\chi$ is defined as : $tr\sqrt{\rho^{1/2}\chi\rho^{1/2}}$. When $\chi = |\psi\rangle\langle\psi|$, $\rho = |\phi\rangle\langle\phi|$ are pure-state, the fidelity reduces to the absolute value of their overlap and equals $|\langle\psi|\phi\rangle|$.

# Chapter 10.

# Discussion and Conclusion

## *Conclusion*

We have studied topologically stable kink solutions in discrete finite models of particles interacting with their nearest neighbours. The spectrum of linear oscillations contains localized gap-separated modes. We found that the nonlinear interactions may be potentially so strong as to defy any harmonic approximation. We have shown also that there exist configurations which allow the benefit of both worlds after quantization – a strongly localized and spectrally isolated mode, yet maintaining a high fidelity for a superposition initial state for at least a couple of tens (almost a hundred) Rabi oscillations.

One hundred periods of coherent oscillation are estimated as sufficient for performing a two-qubit gate with high fidelity in the Cirac-Zoller scheme in the ion trap. In this scheme there are two crucial requirements : the phonon used for QIP must be spectrally isolated and the inter-ion separation must be large enough as to allow single-addressing of individual ions. A high-frequency kink phonon is spectrally separated by a gap and, being localized neither the gap nor the phonon's frequency depend on the trap size. Thus there is no degradation of the gate-speed at large ion numbers. In addition, since it is anti-symmetric, it is expected to couple weakly to external fields. Thus we have shown that, if realized in the ion trap, such a kink phonon makes a natural candidate for solving the scalability problem in the ion trap. Realizing the Frenkel-Kontorova model with trapped ions (without application to QIP) has already been suggested in [23].

We now mention shortly the possibility to transport quantum information using a moving kink. In this work we have explicitly investigated only static kinks, but those can always be Lorentz-boosted to start moving. The localized modes are bound to the kink-core and thus the question arises whether coherent transportation of information is an option. The results of this work suggest that this may not be possible using highly discrete kinks, because of the strength of nonlinearity. Putting the kink into motion would require a large number of low-mode phonons excited, and these will quickly decohere the information stored in the high mode. In a more extended kink, the frequency of the low mode drops quickly and the perturbative treatment would no longer be valid. We may hypothesize that at a large separation of scales between the low and high mode (with a highly extended kink), a regime of parameters may be found where the nonlinear terms are very weak and adiabatic motion of the kink could be possible without significant interaction between the modes, owing to the separation of energies. In a different way, it is perhaps possible that non-perturbative phenomena, such as that displayed by breathers, could manifest an isolated localized mode which has very small dissipative coupling.

## *Numerical and Scalability Considerations*

We conclude with a short discussion of some numerical and physical aspects of our work. First we refer to our choice of configuration. The dominant contribution to the decoherence comes from multi-phonon processes involving the localized high and low modes. Powers of $\omega_2^{-1/2}$ enter as prefactors of the interaction coefficients. Thus keeping the low mode's frequency high is essential in order to minimize the decoherence. A second consideration is the appearance of a resonance $\omega_1 - \omega_2 - \omega_j \to 0$, where $\omega_j$ is any non-localized phonon. Four-phonon resonances are possible similarly. We touch on these points in the following paragraphs, from a slightly more general perspective.

We start by considering the scalability of our results to a large number of particles $N$. In any kink, the local modes do not change their frequency or local interaction strength with increasing $N$ at a given geometry. The interaction of the high mode with all other modes in the chain is given (to the 3$^{rd}$ order) by

$$\frac{1}{3!} \sum_{1jk} \sum_{mns} \frac{\partial^3 V}{\partial x_m \partial x_n \partial x_s} \lambda_{m1} \lambda_{nj} \lambda_{sk} \Theta_1 \Theta_j \Theta_k , \qquad (10.1)$$

where the derivative of the potential is calculated at the equilibrium point of all particles, and $\lambda_{ab}$ is the vector of normal mode $b$, with $a$ the particle index.

In this expression, $\lambda_{m1}$ is the profile of the localized high mode, which has a support on a small and fixed number of particles. Examining the spectrum of the modes, we discover (as mentioned above), that the resonant processes are those with at least one of *j* or *k* (assume it is *j*) equaling *2* (i.e. with the low mode). In this case, $\lambda_{n2}$ has again a compact support. For *k* a plane-wave phonon, the number of resonant processes $\omega_1 + \omega_2 - \omega_k \to 0$ grows proportional to $\sim N$, due to the density of phonon states near $\omega_k \simeq \omega_1 + \omega_2$. However the strength of each process drops with $\lambda_{sk} \sim 1/\sqrt{N}$ (where *s* must be a particle neighbouring the kink particles), to a total expected increase in the decoherence scaling as $\sim \sqrt{N}$. The number of four-phonon resonances on the other hand, is expected to cancel with the vanishing of the local components of the plane-wave normal modes, $\lambda_{sk}\lambda_{ql} \sim \left(1/\sqrt{N}\right)^2$.

These rough arguments suggest that the results we presented may not be scalable to large numbers of particles *N*. However there is a way to circumvent the nonscalability of 3$^{rd}$ order processes, by making them off-resonant. Since the high mode is gap-separated from the phonon band, it is possible to lower the frequency of the low mode to a value which would put all three-phonon processes out of resonance. This would come at the price of increasing the perturbative coefficient standing in front of any term in the Hamiltonian involving the low mode. Another option is to use a different structure of localized modes. As mentioned briefly in Chapter 4, there are discrete models which admit localized modes lying atop the phonon band. The resonance structure of the processes involving such localized modes will necessarily be different. We did not examine these two possibilities in the current work.

Regarding the order of interaction terms retained in the Hamiltonian (8.2), we see numerically that for the sine-Gordon model, the contribution of higher-order terms is negligible in the temperatures we investigate. For the $\phi^4$ model, the geometric coefficients $M_{ijkl}$ are not necessarily decreasing compared with the third order $L_{ijk}$. In this model there are no higher-order terms, but in fact in other models this may be a non-trivial problem. However at small excitations, powers of $\sqrt{\hbar}$ dominate the convergence of this series. In our treatment we kept the quartic terms as a means of validation in the calculation, although their contribution was in fact negligible in the configuration presented.

The calculations in this work were run in Matlab, using code written by the author. The fidelity distance between two density matrices was calculated using the library component QLib [24].

# Appendix A

The master equation as written in equation (8.16) is valid provided that $tr_B\{\tilde{H}_{SB}(t)\chi_B\} = 0$. This is achieved by renormalizing the system Hamiltonian with respect to its mean interaction with the bath, using

$$H_S \rightarrow H_S + tr_B\{H_{SB}\chi_B\}$$
$$H_{SB} \rightarrow H_{SB} - tr_B\{H_{SB}\chi_B\} \qquad (A.1)$$

where :

$$tr_B\{H_{SB}\chi_B\} = \hbar \sum_\alpha s_\alpha \langle B_\alpha \rangle_B. \qquad (A.2)$$

The five bath-operators contribute five different terms

$$\nu_i \equiv \langle B_{ii} \rangle_B = 6 \cdot \frac{1}{4!} \sum_{k \neq 1,2} M_{iikk} \hbar \left(16\omega_i^2 \omega_k^2\right)^{-1/2} (1 + 2n_k)$$

$$\xi_i \equiv \langle B_i \rangle_B = 3 \cdot \frac{1}{3!} \sum_{k \neq 1,2} L_{ikk} \hbar^{1/2} \left(8\omega_i \omega_k^2\right)^{-1/2} (1 + 2n_k)$$

$$\nu_{12} \equiv \langle B_{12} \rangle_B = 6 \cdot \frac{1}{4!} \sum_{k \neq 1,2} M_{12kk} \hbar \left(16\omega_1 \omega_2 \omega_k^2\right)^{-1/2} (1 + 2n_k) \qquad (A.3)$$

These five terms enter the system Hamiltonian as prefactors of the corresponding five system operators, and result respectively in a frequency-shift in each system oscillator, in a mean linear conservative force which is a displacement of each oscillator's 'position', and in a new bilinear interaction created between the two oscillators ;

$$\omega_i \rightarrow \omega_i \left(1 + 2\omega_i^{-2}\nu_i\right)^{1/2}$$
$$\left(a_i^\dagger + a_i\right) \equiv \Theta_i \rightarrow \Theta_i + \omega_i^{-2}\xi_i \Rightarrow H_S \rightarrow H_S - \omega_i^{-2}\xi_i^2/2$$
$$\nu_{12}\left(a_1^\dagger + a_1\right)\left(a_2^\dagger + a_2\right) \qquad (A.4)$$

The frequency appearing in the renormalization of the position operator is the renormalized frequency. We note that the corrections introduced by these terms are all rather small.

The renormalized correlation functions now read

$$\tilde{C}_{ij}(\tau) \equiv \left\langle \left(\tilde{B}_i(t)-\xi_i\right)\left(\tilde{B}_j(t')-\xi_j\right)\right\rangle_B =$$

$$= \frac{\hbar(8\omega_i 8\omega_j)^{-1/2}}{2!2!} tr_B \chi_B \sum_{klmn \neq 1,2} L_{ikl} L_{jmn} (\omega_k \omega_l \omega_m \omega_n)^{-1/2} \left[\left(\tilde{a}_k^\dagger + \tilde{a}_k\right)\left(\tilde{a}_l^\dagger + \tilde{a}_l\right)\right]_t \left[\left(\tilde{a}_m^\dagger + \tilde{a}_m\right)\left(\tilde{a}_n^\dagger + \tilde{a}_n\right)\right]_{t'} - \xi_i \xi_j =$$

$$= \frac{\hbar(8\omega_i 8\omega_j)^{-1/2}}{2!2!} \sum_{klmn \neq 1,2} L_{ikl} L_{jmn} (\omega_k \omega_l \omega_m \omega_n)^{-1/2} \times$$

$$\times \left\{ \begin{aligned} &\delta_{kl}\delta_{mn}(1+2n_k)(1+2n_m) + \delta_{km}\delta_{ln}\left[n_k e^{i\omega_k \tau} + (1+n_k)e^{-i\omega_k \tau}\right]\left[n_l e^{i\omega_l \tau} + (1+n_l)e^{-i\omega_l \tau}\right] + \\ &+ \delta_{kn}\delta_{lm}\left[n_k e^{i\omega_k \tau} + (1+n_k)e^{-i\omega_k \tau}\right]\left[n_l e^{i\omega_l \tau} + (1+n_l)e^{-i\omega_l \tau}\right] \end{aligned} \right\} - \xi_i \xi_j =$$

$$= \frac{\hbar(8\omega_i 8\omega_j)^{-1/2}}{2!2!} \sum_{kl \neq 1,2} (\omega_k \omega_l)^{-1} \times$$

$$\times \left\{ L_{ikk} L_{jll}(1+2n_k)(1+2n_l) + 2L_{ikl} L_{jkl}\left[n_k e^{i\omega_k \tau} + (1+n_k)e^{-i\omega_k \tau}\right]\left[n_l e^{i\omega_l \tau} + (1+n_l)e^{-i\omega_l \tau}\right] \right\} - \xi_i \xi_j =$$

$$= \frac{\hbar(8\omega_i 8\omega_j)^{-1/2}}{2} \sum_{kl \neq 1,2} (\omega_k \omega_l)^{-1} L_{ikl} L_{jkl}\left[n_k e^{i\omega_k \tau} + (1+n_k)e^{-i\omega_k \tau}\right]\left[n_l e^{i\omega_l \tau} + (1+n_l)e^{-i\omega_l \tau}\right] \tag{A.5}$$

$$\tilde{C}_{\alpha\beta\gamma\delta}(\tau) \equiv \left\langle \left(\tilde{B}_{\alpha\beta}(t)-\nu_{\alpha\beta}\right)\left(\tilde{B}_{\gamma\delta}(t')-\nu_{\gamma\delta}\right)\right\rangle_B =$$
$$\alpha,\beta,\gamma,\delta \in \{1,2\}$$

$$= \frac{\hbar}{2!2!} tr_B \chi_B \sum_{kl \neq 1,2} L_{\alpha\beta k} L_{\gamma\delta l}\left(8\omega_\alpha \omega_\beta \omega_k 8\omega_\gamma \omega_\delta \omega_l\right)^{-1/2} \left(\tilde{a}_k^\dagger + \tilde{a}_k\right)\Big|_t \left(\tilde{a}_l^\dagger + \tilde{a}_l\right)\Big|_{t'}$$

$$+ \frac{\hbar^2}{2!2!} tr_B \chi_B \sum_{klmn=1,2} M_{\alpha\beta kl} M_{\gamma\delta mn}\left(16\omega_\alpha \omega_\beta \omega_k \omega_l 16\omega_\gamma \omega_\delta \omega_m \omega_n\right)^{-1/2} \ldots - \nu_{\alpha\beta}\nu_{\gamma\delta} =$$

$$= \frac{\hbar}{2!2!} \sum_{k \neq 1,2} L_{\alpha\beta k} L_{\gamma\delta k}\left(8\omega_\alpha \omega_\beta 8\omega_\gamma \omega_\delta \omega_k^2\right)^{-1/2}\left[n_k e^{i\omega_k \tau} + (1+n_k)e^{-i\omega_k \tau}\right]$$

$$+ \frac{\hbar^2}{2!2!} \sum_{kl \neq 1,2}\left(16\omega_\alpha \omega_\beta 16\omega_\gamma \omega_\delta \omega_k^2 \omega_l^2\right)^{-1/2} \times$$

$$\times \left\{ M_{\alpha\beta kk} M_{\gamma\delta ll}(1+2n_k)(1+2n_l) + 2M_{\alpha\beta kl} M_{\gamma\delta kl}\left[n_k e^{i\omega_k \tau} + (1+n_k)e^{-i\omega_k \tau}\right]\left[n_l e^{i\omega_l \tau} + (1+n_l)e^{-i\omega_l \tau}\right] \right\} - \nu_{\alpha\beta}\nu_{\gamma\delta} =$$

$$= \frac{\hbar}{2!2!} \sum_{k \neq 1,2} L_{\alpha\beta k} L_{\gamma\delta k}\left(8\omega_\alpha \omega_\beta 8\omega_\gamma \omega_\delta \omega_k^2\right)^{-1/2}\left[n_k e^{i\omega_k \tau} + (1+n_k)e^{-i\omega_k \tau}\right]$$

$$+ \frac{\hbar^2}{2} \sum_{kl \neq 1,2}\left(16\omega_\alpha \omega_\beta 16\omega_\gamma \omega_\delta \omega_k^2 \omega_l^2\right)^{-1/2} M_{\alpha\beta kl} M_{\gamma\delta kl}\left[n_k e^{i\omega_k \tau} + (1+n_k)e^{-i\omega_k \tau}\right]\left[n_l e^{i\omega_l \tau} + (1+n_l)e^{-i\omega_l \tau}\right] \tag{A.6}$$

$$\tilde{C}_{i\gamma\delta}(\tau) \equiv \left\langle \left(\tilde{B}_i(t) - \xi_i\right)\left(\tilde{B}_{\gamma\delta}(t') - \nu_{\gamma\delta}\right)\right\rangle_B =$$
$$i,\gamma,\delta \in \{1,2\}$$

$$= \frac{\hbar^{3/2}}{2!2!} tr_B \chi_B \sum_{jkmn=1,2} L_{ijk} M_{\gamma\delta mn} \left(8\omega_i\omega_j\omega_k 16\omega_\gamma\omega_\delta\omega_m\omega_n\right)^{-1/2} \ldots - \xi_i\nu_{\gamma\delta} =$$

$$= \frac{\hbar^{3/2}}{2!2!} \sum_{km \neq 1,2} \left(8\omega_i\omega_k^2 16\omega_\gamma\omega_\delta\omega_m^2\right)^{-1/2} \times$$

$$\times \left\{ \begin{array}{l} L_{ikk} M_{\gamma\delta mm}(1+2n_k)(1+2n_m) + \\ 2L_{ikm} M_{\gamma\delta km}\left[n_k e^{i\omega_k\tau} + (1+n_k)e^{-i\omega_k\tau}\right]\left[n_m e^{i\omega_m\tau} + (1+n_m)e^{-i\omega_m\tau}\right] \end{array} \right\} - \xi_i\nu_{\gamma\delta} =$$

$$= \frac{\hbar^{3/2}}{2} \sum_{km \neq 1,2} \left(8\omega_i\omega_k^2 16\omega_\gamma\omega_\delta\omega_m^2\right)^{-1/2} L_{ikm} M_{\gamma\delta km}\left[n_k e^{i\omega_k\tau} + (1+n_k)e^{-i\omega_k\tau}\right]\left[n_m e^{i\omega_m\tau} + (1+n_m)e^{-i\omega_m\tau}\right]$$

(A.7)